\documentstyle[prd,eqsecnum,aps,psfig]{revtex}
\date{2 February 1998}
\begin{document}

%
%
%
%
\def\ptitle#1{\wideabs{\maketitle
\abstract #1 \endabstract \pacs{Pacs numbers: 03.70.+k, 98.80.Cq}}}

\draft

\title{The late-time singularity inside non-spherical black holes}
\author{Patrick R. Brady}
\address{Theoretical Astrophysics  130-33,  California Institute of
	Technology,  Pasadena, CA 91125 }
\author{Serge Droz}
\address{Department of Physics,  University of Guelph,  N1G 2W1 Canada}
\author{Sharon M. Morsink}
\address{Department of Physics,  University of Wisconsin-Milwaukee, P.O.
Box
413, Milwaukee, WI 53201}
\preprint{GRP-491}

\typeout{ABSTRACT}

\ptitle{
It was long believed that the singularity inside a realistic, rotating
black hole must be spacelike.  However, studies of the internal
geometry of black holes indicate a more complicated structure is
typical.  While it seems likely that an observer falling into a 
black hole with the
collapsing star encounters a crushing spacelike singularity, an
observer falling in at late times generally reaches a null singularity
which is vastly different in character to the standard Belinsky,
Khalatnikov and Lifschitz (BKL) spacelike
singularity~\mbox{[}V.~A. Belinsky, I.~M. Khalatnikov, and
E.~M. Lifshitz, Sov. Phys. JETP {\bf 32}, 169 (1970)\mbox{]}.  In the
spirit of the classic work of BKL we present an asymptotic analysis of
the null singularity inside a realistic black hole.  Motivated by
current understanding of spherical models, we argue that the Einstein
equations reduce to a simple form in the neighborhood of the null
singularity.  The main results arising from this approach are
demonstrated using an almost plane symmetric model.  The analysis
shows that the null singularity results from the blueshift
of the late-time gravitational wave tail; the
amplitude of these gravitational waves is taken to decay as an inverse
power of advanced time as suggested by perturbation theory.  The 
divergence of the Weyl curvature at the null
singularity is dominated by the propagating modes of the gravitational
field, that is $$ C_{\alpha\beta\gamma\delta}
C^{\alpha\beta\gamma\delta} \sim
\Psi_0\Psi_4 \sim v^{-(2l+3)} e^{2 \kappa v} 
$$ 
as $v\rightarrow \infty$ at the Cauchy horizon.  Here, $\Psi_0$ and
$\Psi_4$ are the Newman-Penrose Weyl scalars, and $l\geq 2$ is the
multipole order of the perturbations crossing the event horizon.  The
null singularity is weak in the sense that tidal distortion remains
bounded along timelike geodesics crossing the Cauchy horizon.  
{These results are in agreement with previous analyses of black
hole interiors.}  We briefly discuss some outstanding problems which 
must be resolved before
the picture of the generic black hole interior is complete.
}

\narrowtext
\clearpage
%
%
%
%
\section{Introduction}

Spacetime singularities are an inevitable consequence of the Einstein
field equations.  They mark the boundary of spacetime, and the limit
of our current understanding of gravitational physics.  Unfortunately,
the powerful techniques used to demonstrate the existence of
singularities say nothing about the character of these physical
blemishes~\cite{Hawking_SW:1973}.

The classic works of Belinsky, Khalatnikov and
Lifschitz~\cite{BKL:1970} address this deficiency by integrating
Einstein's equations in the neighborhood of a spacelike singularity.
They present compelling evidence that the general solution takes an
inhomogeneous Kasner form exhibiting chaotic oscillations of the
Kasner axes as a crushing singularity is approached~\cite{BKL:1970}.
The functional genericity of the BKL solutions is widely interpreted
as indicating that all physical singularities must be of this form.

Nonetheless, studies of the internal geometry of black holes suggest a
picture involving two distinct regimes. Observers falling into the
black hole with the collapsing star generally encounter a spacelike
singularity (which is presumably of the BKL type).  On the other hand,
observers that fall in at late times, when the external geometry has
settled down to an almost stationary state, encounter a weak, null
singularity\footnote{The weakness of the mass-inflation singularity
was first elucidated by Ori in Refs.~\cite{Ori_A:1991,Ori_A:1992}} 
of a type similar to the mass-inflation singularity of
Poisson and Israel~\cite{Poisson_E:1990}.

To understand this behavior we must first discuss the formation of a
black hole by the gravitational collapse of a rotating star.  At late
times, the external gravitational field is believed to settle down to
a Kerr-Newman solution.  These solutions have a timelike singularity
which is preceded by a Cauchy horizon---a null hypersurface marking
the boundary of the domain of dependence for Cauchy data prescribed in
the black hole exterior.  The Cauchy horizon is non-singular, and the
spacetime can be analytically extended through it.  The global
solution then suggests that black holes act as tunnels from our
asymptotically flat universe to other identical, but distinct,
universes.  Indeed there is an infinite lattice of universes extending
into the past and future of our own.  Furthermore, observers may
travel through the ring singularity inside these black holes, passing
to achronal regions of spacetime.

In the late 1960's Penrose pointed out that the Cauchy horizon inside
such a black hole is unstable~\cite{Penrose_R:1969}.  Time-dependent
perturbations originating outside the black hole get infinitely
blueshifted as they propagate inwards near to the Cauchy horizon,
consequently, the energy density associated with these perturbations
diverges as measured by a free-falling observer attempting to cross
through the horizon.  Perturbative calculations~\cite{Chandra_S:1982}
in Reissner-Nordstr\"om and Kerr spacetimes have validated Penrose's
original arguments.

In general, gravitational collapse is expected to be asymmetric, so
that gravitational and electro-magnetic waves are emitted by a newly
formed black hole as it settles down to a stationary, axisymmetric
state.  Detailed studies of perturbations of black hole geometries
show that such gravitational wave emission results in wave tails which
decay according to an inverse power law of time in the exterior of the
black hole~\cite{Price_R:1972,Gundlach_C:1994}.  Some of this
radiation inevitably crosses the event horizon getting infinitely
blueshifted near to the Cauchy horizon.  For this reason one might
expect the internal geometry of a black hole formed by collapse is
significantly different to that of the exact stationary solutions.

\subsection{Spherical models of black hole interiors}

The first attempts to understand the back-reaction of the blueshifted,
radiative tail on the internal geometry of black holes were restricted
to spherical symmetry.  Hiscock~\cite{Hiscock_W:1981} argued that
Isaacson's~\cite{Isaacson_R:1968} effective stress-energy description
of high-frequency gravitational waves should be valid near the Cauchy
horizon.  He considered a charged black hole with a directed influx of
lightlike dust with stress-energy tensor $T_{\alpha\beta}=\rho_{\rm
in} \ell_\alpha\ell_\beta$ where $\ell_\alpha\ell^\alpha=0$, and
showed that an observer dependent singularity forms along the Cauchy
horizon in this circumstance.

In reality, some of the infalling radiation is back-scattered off the
curvature inside the event horizon.  Poisson and
Israel~\cite{Poisson_E:1990} modeled this effect by another flux of
lightlike dust moving to the right; they demonstrated, for the first
time, that non-linear effects transform the Cauchy horizon into a
scalar curvature singularity.  It is worth summarizing the essence of
their argument here.

\begin{figure}
\centerline{
\psfig{file=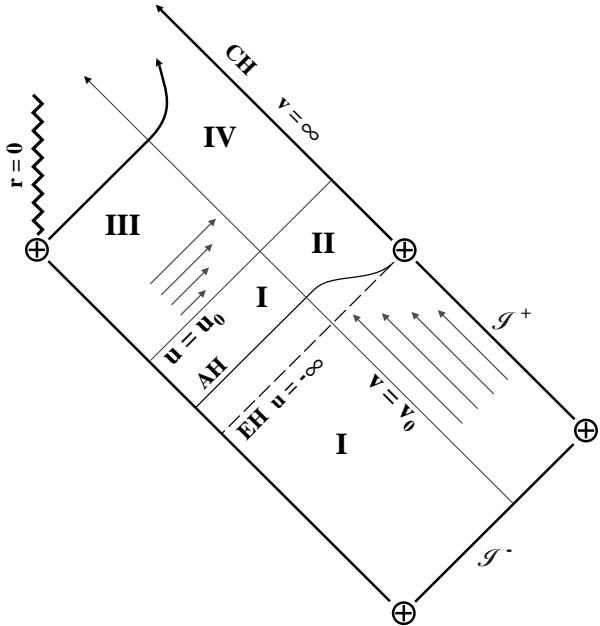,width=8cm,bbllx=9pt,bblly=326pt,bburx=489pt,bbury=827pt}
}
\caption{\label{fig:PImodel} The Poisson-Israel spherical model of the
black hole interior.  The infalling wave-tail of gravitational
radiation, and its scattered component, are modeled by lightlike dust.
The region of purely ingoing (outgoing) dust is indicated by II (III)
in the figure, and is described by a charged Vaidya solution.  The
solutions in regions II and III are matched to a Reissner-Nordstr\"om
solution (region I) along the null rays $v=v_0$ and $u=u_0$. As usual
${\cal I}^+$ and ${\cal I}^-$ denote future and past null infinity.
The inflow causes the apparent horizon (AH) to expand and smoothly
approach the event horizon (EH) of the final black hole.  The outflow
in region IV causes the, initially static, Cauchy horizon (CH) to
contract. This, together with the fatal blue shift experienced by the
inflow, causes a curvature singularity to develop along CH.  }
\end{figure}

Figure~\ref{fig:PImodel} shows the setting for the characteristic
initial value problem.  Charged ingoing and outgoing Vaidya solutions
in regions II and III, respectively, are matched continuously onto
region I, which is described by a static Reissner-Nordstr\"om solution
with mass $m_0$ and charge $q$.  (The pure ingoing region, II, was
studied by Hiscock.)  In region IV the line element can be written as
\begin{equation}
	ds^2= -2e^\lambda du dv + r^2 (d\theta^2 + \sin^2\theta
		d\phi^2)\; , \label{eq:1.1}
\end{equation}
where $r=r(u,v)$ and $\lambda=\lambda(u,v)$.  The coordinates are
chosen so that $v$ is standard advanced time, {\it i.e.} $v=\infty$ at
future null infinity, and the retarded time $u=-\infty$ on the black
hole event horizon.  The stress-energy of cross-flowing null dust is
\begin{equation}
	T_{\alpha\beta} = \frac{L_{\rm in}(v)}{4\pi r^2} \ell_\alpha
 		\ell_\beta + \frac{L_{\rm out}(u)}{4\pi r^2} n_\alpha
 		n_\beta\; , \label{eq:1.2}
\end{equation}
where $\ell_\alpha= -\partial_\alpha v$ and $n_\alpha=-\partial_\alpha
u$.  The luminosity function $L_{\rm in}(v)$ is fixed by requiring that
the flux of stress-energy across the event horizon decays as an
inverse power of advanced time, thus
\begin{equation}
	L_{in}(v) = \alpha (\kappa v)^{-q}\; , \label{eq:1.3}
\end{equation}
with $q = 4l+6$, where $l$ is the multipole order of the perturbing
field, fixed by Price's analysis~\cite{Price_R:1972}.  The constant
$\alpha$ depends on the luminosity of the star that collapses to form
the black hole, and $\kappa$ is taken to be the surface gravity of the
stationary segment of the Cauchy horizon in region II.  This
functional form is motivated by our understanding of radiative tails
in the exterior of the black hole~\cite{Price_R:1972}.  The outflux is
produced by scattering of ingoing tail radiation inside the black
hole, and consequently also has a power law
form~\cite{Chandra_S:1982,Gursel_Y:1979,Ori_A:1997a,Bonanno_A:1994}
\begin{equation}
	 L_{\rm out}(u) = \beta (-\kappa u)^{-q} \;   \label{eq:1.4}
\end{equation}
for large negative values of the coordinate $u$.  

In the limit as $v\rightarrow \infty$, the corresponding solution of
Einstein's equations is well approximated by
\begin{eqnarray}
	e^{\lambda} &\simeq& \frac{e^{-\kappa v}}{r} \label{eq:1.5}\\
	r^2 &\simeq& r_{\rm CH}^2 +\frac{2\alpha (\kappa v)^{-q+1}}{
		\kappa^2 (q-1)} - \frac{2\beta  (-\kappa u)^{-q+1}}{
		\kappa^2 (p-1)} \; , \label{eq:1.6}
\end{eqnarray}
where $r_{\rm CH}$ is the constant radius of the Cauchy horizon in region II.  The
square of the Weyl tensor diverges on the Cauchy horizon in this
solution, yet the radial function given by Eq.~(\ref{eq:1.6}) is
non-zero for sufficiently large $|u|$.  This indicates that the
singularity is not a central ($r=0$) singularity with which we are
familiar, in fact it is a {\em null} singularity.  Spherical symmetry
allows the introduction of a mass function $m(u,v)$ which is directly
related to the Weyl curvature of the spacetime:
\begin{eqnarray}
	m(u,v) &&= \left(\mbox{$\textstyle \frac{1}{2}$} C{^{\theta\phi}}_{\theta\phi} r^3 +
		q^2/r \right) \label{eq:1.7} \\
		&&\stackrel{v\rightarrow\infty}{\longrightarrow}
		\frac{\alpha\beta e^{\kappa v}}{\kappa r^2}
		(\kappa v)^{-q}(-\kappa u)^{-q}  \; . \label{eq:1.8}
\end{eqnarray}
The divergence of this mass function as $v\rightarrow\infty$ in region
IV prompted Poisson and Israel to refer to the accompanying
scalar-curvature singularity as a {\em mass inflation} singularity.

Subsequently, Ori~\cite{Ori_A:1991} pointed out that physical objects
which encounter a mass-inflation singularity experience only finite
tidal distortion, even though the tidal forces acting on the object
diverge.  In this sense, the mass-inflation singularities are weak
although the relevance of this point continues to be
debated~\cite{Balbinot_R:1991,Hermann_R:1992}.

A variety of spherical models have now been
studied~\cite{Gnedin:1993,Brady_P:1995a,Burko_L:1997}, all of them
indicate the presence of a null, scalar curvature singularity at the
Cauchy horizon.  The numerical analyses in \cite{Brady_P:1995a} also
demonstrates the slow contraction of the null generators of the Cauchy
horizon to zero radius, and the formation of a spacelike singularity
deep inside the black hole core.

\subsection{Beyond spherical symmetry}

While simplified models argue strongly in favor of weak, null
singularities inside black holes, the results are restricted entirely
to spherical symmetry.  Nevertheless the physics behind the
mass-inflation phenomenon is extremely general.  Perturbations
originating in the external universe get infinitely blueshifted as
they propagate close to the Cauchy horizon resulting in a scalar curvature
singularity through non-linear interaction with gravity.  Attempts
have been made to investigate the problem in less symmetric
situations.  Ori~\cite{Ori_A:1992} has used non-linear perturbation theory
to examine back-reaction of perturbations on the Kerr geometry.  His
results support the existence of a weak, null singularity in this
case.  Bonanno~\cite{Bonanno_A:1996} matched two radiating Kerr solutions
along a null hypersurface and showed, in the limit of small angular
momentum, that the mass-function diverges on a null hypersurface.

Brady and Chambers~\cite{Brady_P:1995b}
demonstrated that weak, null singularities are consistent with the Einstein
equations on a pair of intersecting null surfaces, and have
investigated the rate of divergence of curvature on a null surface
crossing the singularity.  An important step has also been taken by
Ori and Flanagan~\cite{Ori_A:1996} who argue that the vacuum, Einstein
equations admit functionally generic solutions containing weak, null
singularities.  Their results invalidate local
arguments~\cite{Yurtsever_U:1993} that null singularities are always
unstable to transformation into crushing spacelike singularities of
the BKL type.

\subsection{Discussion and overview of this paper}

In this paper we demonstrate that a null singularity replaces the
Cauchy horizon inside a black hole formed by gravitational collapse.
The method employed in the following analysis is similar in spirit to
that of BKL~\cite{BKL:1970}; we use an asymptotic expansion to find
the leading order behavior of the metric near the Cauchy horizon
without making any assumptions about the symmetry of the spacetime.
In this way we show that the wave tail of gravitational collapse
results in the formation of a null curvature singularity at the Cauchy
horizon.

{Our results should be compared with those obtained using
non-linear perturbation theory.  In that context,
Ori~\cite{Ori_A:1992} has previously argued that a weak, null
singularity is present along the Cauchy horizon of a generic, rotating
black hole.  Our analysis, being asymptotic, relies on physical
arguments to provide initial data near to the Cauchy horizon; in this
sense, it is a local analysis.  Nevertheless, the results presented
below embody all orders of perturbation theory and lend further
support to the claim that the perturbative approach captures the
essential features of spacetime structure near the Cauchy horizon
singularity inside a realistic, rotating black hole~\cite{Ori_A:1992}.}

It is natural to use null coordinates to attack this problem since the
Cauchy horizon is a null hypersurface.  To facilitate the analysis we
employ a double-null formalism \cite{Brady_P:1996} in which spacetime
is decomposed into two families of intersecting null hypersurfaces.
This decomposition of the Einstein field equations is reviewed in
Sec.~\ref{sec:double-null} where it is shown that the most general
double-null metric depends on six functions of four variables.
Interested readers are referred to Brady {\em et
al}~\cite{Brady_P:1996} for more details.

The main assumption made in this paper is that the results of
scattering on the Kerr background~\cite{Ori_A:1997b} are sufficient to
determine the initial data for the interior problem. This assumption
and other physical considerations are discussed in
Sec.~\ref{sec:assumptions}.

In Sec.~\ref{sec:simple} we present the essential features of our
arguments in a simplified context where the gravitational field has
only one dynamical degree of freedom.  We carefully construct the
solution in the neighborhood of the Cauchy horizon for this model problem.
Moreover, we demonstrate that the solution is characterized by a
singularity at which the Weyl curvature diverges like
\begin{equation}
C_{\alpha\beta\gamma\delta} C^{\alpha\beta\gamma\delta} \sim v^{-(2l+3)}
e^{2 \kappa v} \ \ \
\end{equation}
as $v \rightarrow \infty$ at the Cauchy horizon.   We show that the
resulting spacetime is not algebraically special except at the
singularity, where it becomes asymptotically Petrov type N.

The relevance of the simplified model is made clear in Sec.
\ref{sec:general-case} where we present the general analysis.  By
neglecting terms in the field equations which we expect to be
exponentially small near to the Cauchy horizon, it is shown that the
general solution is identical in character to that presented in
Sec. \ref{sec:simple}.  

Finally, we discuss the strength of the singularity in
Sec. \ref{sec:strength}. While tidal forces experienced by an observer
diverge at the singularity, the integrated tidal distortion is finite.
{As discussed by Ori~\cite{Ori_A:1991,Ori_A:1992},  the singularity is
therefore weak and we are once again faced with the possibility that
spacetime might extend beyond the Cauchy horizon (in a non-analytic
manner).}  We emphasize that gravitational shock waves are the only
type of curvature which can be confined to a thin layer in the absence
of matter~\cite{Balbinot_R:1991,Barrabes_C:1991}.  This suggests that
a classical continuation of spacetime beyond the Cauchy horizon is
unlikely.

We adopt Misner, Thorne and Wheeler curvature conventions~\cite{MTW}
with metric signature $(-+++)$ throughout the text.  Greek indices
$\alpha,\beta,\dots$ run from 0 to 3; upper-case Latin indices
$A,B,\dots$ take values $(0,1)$; and lower-case Latin indices
$a,b,\dots$ take values $(2,3)$.

%
%
%
%

\section{Double null formalism}\label{sec:double-null}
The complexity of the Einstein equations requires the careful
selection of a formalism with which to pursue our goal of
understanding the nature of spacetime near null singularities.
In~\cite{Brady_P:1996} we have developed the necessary machinery based
on a dual null decomposition of the Einstein field equations.  In this
section we present those details which are necessary for the
application at hand.  The interested reader is referred
to~\cite{Brady_P:1996} for a more complete treatment.

\subsection{Lightlike foliation of spacetime}

We suppose that we are given a foliation of spacetime by lightlike
hypersurfaces $\Sigma^0$ with normal generators $\ell^{(0)}$, and a
second, independent foliation by lightlike hypersurfaces $\Sigma^1$
with generators $\ell^{(1)}$ nowhere parallel to $\ell^{(0)}$. The
intersections of $\{\Sigma^0\}$ and $\{\Sigma^1\}$ define a foliation
of codimension 2 by spacelike 2-surfaces $S$.  (The topology of $S$ is
unspecified. All our considerations are local.)  $S$ has exactly two
lightlike normals at each of its points, co-directed with $\ell^{(0)}$
and $\ell^{(1)}$.

In terms of local charts, the foliation is described by the embedding
relations 
\begin{equation} 
	x^\alpha=x^\alpha(u^A,\theta^a).
	\label{eq:1} 
\end{equation} 
Here, $x^\alpha$ are four-dimensional spacetime co-ordinates; $u^0= v$
and $u^1=u$ are a pair of scalar fields constant over each of the
hypersurfaces $\Sigma^0$ and $\Sigma^1$ respectively.  The intrinsic
co-ordinates $\theta^a$ of the 2-spaces $S$, each characterized by a
fixed pair of values $(u^0,u^1)$, are convected along the vectors
$\partial/\partial u^A$.

It is convenient to effect a partial normalization of the lightlike
vectors $\ell^{(A)}$ by imposing the condition
\begin{equation}
	{}^{(4)}g^{\alpha\beta}
	\ell^{(A)}_\alpha \ell^{(B)}_\beta = e^\lambda \eta^{AB}\equiv
		e^{\lambda} \, \mbox{\rm anti-diag}\, (-1,-1)
\end{equation}
for some scalar field $\lambda(x^\alpha)$, reflecting the freedom to
arbitrarily rescale a null vector.  The superscript $(4)$ is used to
indicate four dimensional objects whenever confusion might arise.  We
can use $\eta^{AB}$ and its inverse $\eta_{AB}$ to raise and lower
uppercase Latin indices.  Furthermore, the vectors $\ell^{(A)}$ are
parallel to the gradients of $u^A$ allowing us to write
\begin{equation}
	\ell^{(A)}_{\alpha}
	= e^\lambda \partial_\alpha u^A \; .
\end{equation}

The pair of vectors $e_{(a)}^\alpha$, defined from (\ref{eq:1}) by
\begin{equation}
	e_{(a)}^\alpha=\partial x^\alpha/\partial\theta^a,
	\label{eq:6}
\end{equation}
are holonomic basis vectors tangent to $S$. The intrinsic metric
$g_{ab}$ of $S$ is determined by their scalar products:
\begin{equation}
	g_{ab}={}^{(4)}g_{\alpha\beta} e_{(a)}^{\alpha}
	e_{(b)}^{\beta}.  \label{eq:7}
\end{equation}
Lower-case Latin indices are lowered and raised with $g_{ab}$ and its
inverse $g^{ab}$.  Since $\ell^{(A)}$ is normal to every
vector in $\Sigma ^A$, we have
\begin{equation}
	\ell^{(A)}_\alpha e_{(a)}^\alpha =0.  \label{eq:8}
\end{equation}

Finally, introducing a pair of shift vectors $s_A^\alpha$ tangent to
$S$ by
\begin{equation} \frac{\partial x^\alpha}{\partial
u^A}=\ell_{(A)}^\alpha+s_A^a \,e_{(a)}^\alpha \; , \label{eq:9}
\end{equation}
an arbitrary displacement $dx^\alpha$ in spacetime is
\begin{equation}
	dx^\alpha = \ell^\alpha_{(A)} du^A + e^\alpha_{(a)} (d\theta^a
		+s^a_{A} du^A) \; .
\end{equation}
Thus, the final form of the line element used below to discuss the
black hole interior is
\begin{equation}
	ds^2=e^\lambda\eta_{AB}
	\,du^A\,du^B+g_{ab}(d\theta^a+s_A^a\,du^A)
	(d\theta^b+s_B^b\,du^B).
	\label{eq:line-element}
\end{equation}
We will impose further coordinate conditions as the need arises, but
it is manifest that Eq.~(2.9) is sufficient to describe the most
general spacetime admitted as a solution to the Einstein field
equations.

\subsection{2+2 decomposition of curvature}

Following~\cite{Brady_P:1996}, the components of the Ricci tensor can
be compactly expressed in terms of two dimensionally covariant
quantities.  In passing to these expression we must first introduce
some notation.  The extrinsic curvatures of $S$ are defined by
\begin{eqnarray}
	2 K_{Aab} &\equiv& e^\alpha_{(a)} e^\beta_{(b)} \left[
	{^{(4)}\mbox{$\cal L$}_{\ell_{(A)}}} \bar{g}_{\alpha\beta}   \right]
	\nonumber \\
	&=& D_A  g_{ab}\;
	,
\end{eqnarray}
where $\bar{g}_{\alpha\beta} = {}^{(4)}g_{\alpha\beta} - e^\lambda
\eta_{AB}\ell^{(A)}_{\alpha}\ell^{(B)}_{\beta} =
e^{(a)}_{\alpha}e^{(b)}_\beta g_{ab}$ is the spatial projection
tensor, and ${^{(4)}\mbox{$\cal L$}_{\ell_{(A)}}}$ is the four
dimensional Lie derivative along $\ell_{(A)}$. We have also introduced
the derivative operators $D_A$ acting on two-dimensional tensorial
objects according to the rule
\begin{equation}
	D_A X{^{a\ldots}}_{b\ldots}
	= (\partial_A - \mbox{$\cal L$}_{s^d_{A}}) X{^{a\ldots}}_{b\ldots} \; .
\end{equation}
Here $\mbox{$\cal L$}_{s^d_{A}}$ is the two-dimensional Lie
derivative.  (Generally $D_A$ can be defined in a four-dimensionally
covariant manner by its action on spatial four
tensors~\cite{Brady_P:1996};
\begin{equation}
	D_A X{^{a\ldots}}_{b\ldots} \equiv e^{(a)}_\alpha e_{(b)}^\beta
	{^{(4)}\mbox{$\cal L$}_{\ell_{(A)}}} ( X{^{c\ldots}}_{d\ldots}
e^\alpha_{(c)}
	e^{(d)}_{\beta} ) \; ,
\end{equation}
thus providing a complete geometrical interpretation of $D_A$ in terms
of Lie derivatives.)

The geometric meaning of the extrinsic curvature is illustrated by Lie
propagating a circle on $S$ along $\ell_{(A)}$.  The expansion rate of
the light rays is given by the trace of the extrinsic curvature $K_A
\equiv g^{ab} K_{Aab}$.  The traceless part of the extrinsic
curvature, $\sigma_{Aab} \equiv K_{Aab} - \mbox{$\textstyle \frac{1}{2}$} g_{ab} K_A$, is the
shear of the null congruence, and the twist is
\begin{eqnarray}
	\omega^\alpha
	&\equiv& [\ell_{(1)}, \ell_{(0)}]^\alpha \nonumber\\
	&=& e^\alpha_{(a)} \epsilon^{AB} (\partial_B s^a_A - s^b_B
	\partial_b s^a_A )
        \label{eq:2.13}
\end{eqnarray}
where $\epsilon_{AB}$ is the anti-symmetric tensor density
($\epsilon_{01} =1$).  Clearly $\omega^\alpha$ is purely spatial.

Finally, the expressions for the Ricci tensor are
\twocolumn[
\widetext {  
\begin{eqnarray} 
^{(4)}\! R_{ab}=R_{\alpha\beta}e_{(a)}^\alpha e_{(b)}^\beta 
	&=& -\mbox{$\textstyle \frac{1}{2}$} g_{ab} e^{-\lambda}(D_AK^A + K_AK^A) + 
	e^{-\lambda}(-D^A\sigma_{Aab} + 2 \sigma{_{A(a}}^{d} 
	\sigma{^A}_{b)d}) \nonumber \\ 
	& & + \frac12\,^{(2)}\!R g_{ab} -\frac12
	e^{-2\lambda}\,{\omega_a\omega_b}-\lambda_{;ab}-
	\frac12\lambda_{,a}\lambda_{,b} \label{r1}\\ 
R_{AB}=R_{\alpha\beta} \ell_{(A)}^\alpha \ell_{(B)}^\beta 
	&=& -D_{(A}K_{B)} - \mbox{$\textstyle \frac{1}{2}$} K_A K_B - \sigma_{Aab} \sigma_B\,^{ab}
	+K_{(A}D_{B)}\lambda \nonumber\\ 
	& &\qquad -\frac12\eta_{AB} \left[
	\left(D^E+K^E\right)D_E\lambda-e^{-\lambda}
	\omega^a\omega_a+(e^\lambda){^{;a}}_{;a} \right] 
	\label{r2}\\ 
R_{Aa}=R_{\alpha\beta} e_{(a)}^\alpha \ell_{(A)}^\beta
	&=&\sigma{{_{Aa}}^{b}}_{;b} - \mbox{$\textstyle \frac{1}{2}$}\partial_a
	K_A-\frac12\partial_aD_A\lambda+\frac12K_A\partial_a\lambda
	\nonumber\\ 
& &\qquad +\frac12 \epsilon_{AB}e^{-\lambda} \left[
	\left(D^B+K^B\right)\omega_a-\omega_a D^B \lambda \right], 
	\label{r3}
\end{eqnarray} 
}]
\narrowtext
where $^{(2)}\!R$ is the curvature scalar associated with
the two-metric $g_{ab}$,  and a semi-colon indicates the
two-dimensional covariant derivative.

\section{Assumptions}
\label{sec:assumptions}
The mathematical theory of black holes is well
established~\cite{Chandra_S:1983}.  Indeed, it is widely
accepted that the external geometry of a black hole formed by the
gravitational collapse of a star is described by a Kerr-Newman
solution at late times.  The mechanism by which the black hole settles
down to this stationary state was elucidated by
Price~\cite{Price_R:1972} for nearly spherical, uncharged collapse,
and his results have been extended to other situations by several
authors~\cite{Gundlach_C:1994,Krivan_W:1996}.  These works provide the
initial conditions for the interior problem by determining the
behavior of the radiative tail of gravitational waves at the event horizon of the
black hole: in particular, the amplitude of the gravitational wave
flux generally decays as an inverse power of advanced time.

Some simplifying assumptions are required in order to make progress
against the black-hole interior problem.  We clarify our approach
below.

\subsection{Physical considerations}\label{subsec:physical}

\begin{figure}
\psfig{file=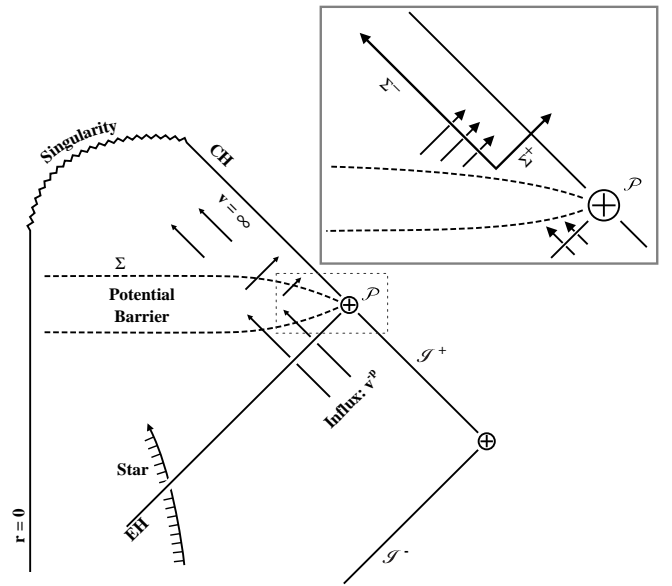,width=7cm,bbllx=0pt,bblly=100pt,bburx=555pt,bbury=757pt}
\caption{\label{fig:generic-spacetime} A schematic representation of
the spacetime structure of a realistic black hole formed by the
collapse of a rotating star.  An observer falling in with the surface
of the star encounters a spacelike singularity indicated by the
zig-zag line in the figure.  Gravitational radiation, originating from
the stellar collapse, gets partially scattered into the black hole; at
late times, this flux decays as an inverse power of external advanced
time $v^{-q}$.  The radiation which crosses the event horizon (EH)
gets scattered once again by the inner gravitational potential barrier
(indicated in the figure).  The result is a flux of gravitational
waves irradiating the Cauchy horizon (CH) and causing it to contract.
Note that CH is located at $v=\infty$ in these coordinates.  The
inset shows the set up of the characteristic initial value problem
solved in the text.  The two initial characteristics $\Sigma^+$:
$u=u_0$ and $\Sigma^-$: $v=v_0$ are indicated.  The shear along these
characteristic surfaces has an inverse power-law dependence on the
advanced and retarded times reflecting the behavior of the late time
wave tail of gravitational collapse.}
\end{figure}

The inner Cauchy horizon of a stationary black hole is unstable to
linear perturbations that undergo infinite gravitational blueshift
there.  The central assumption of our analysis is that an asymptotic
limit exists in which effects on the geometry, inside a black hole
formed by gravitational collapse, are dominated by gravitationally
blueshifted tail-radiation which propagates into the black hole at
late times (as measured by external observers).  The postulated
structure of spacetime inside the black hole is shown in
Fig.~\ref{fig:generic-spacetime}.  On physical
grounds~\cite{Israel_W:1991} one expects that the region between the
event horizon and the three dimensional spacelike surface {$\Sigma$}
can be adequately described by linear perturbation theory---there is
no physical mechanism to excite an instability of the interior before
the blueshifting takes hold.  We examine the structure of spacetime in
the region to the future of {$\Sigma$} and near to the Cauchy horizon,
given initial data consistent with scattering of a test gravitational
wave field inside a Kerr black hole~\cite{Ori_A:1997b}.

The late-time wave tail of gravitational collapse produces a flux of
radiation across the black-hole event horizon which decays as an
inverse power-law of external advanced time~\cite{Krivan_W:1996}.
Starting from these initial conditions Ori~\cite{Ori_A:1997b} has
examined the scattering of a test field inside a Kerr black hole.  His
results indicate that the amplitudes of test fields decay as an
inverse power-law in both retarded and advanced times $u$ and $v$ near
to $\cal P$ in Fig.~\ref{fig:generic-spacetime}.  Ori also argues that
the decay of non-axisymmetric modes is modulated by oscillatory terms
which originate from the rotation of the inner horizon with respect to
infinity, {\it i.e.}  the oscillations are a direct consequence of
frame dragging in the Kerr geometry.  Since these oscillations have no
counterpart in spherical symmetry it seems worthwhile to outline Ori's
argument here.

The line element for the Kerr spacetime,  written in familiar
Boyer-Lindquist coordinates,  is 
\begin{eqnarray}
 	ds^2 &=& - (1-2Mr/\rho^2) dt^2 + (\rho^2/\Delta) dr^2 + \rho^2
 	d\theta^2 \nonumber \\
	&& + [r^2+a^2+(2Mra^2/\rho^2)\sin^2\theta] \sin^2\theta
	d\phi^2 \nonumber \\ 
	&&- (4Mra/\rho^2)\sin^2\theta d\phi dt \; , \label{eq:kerr}
\end{eqnarray}
where $\rho^2 = r^2 + a^2\cos^2\theta$ and $\Delta = r^2-2Mr+a^2$.
The solution depends on two parameters: the mass $M$, and the angular
momentum per unit mass $a$.  The equations governing perturbations of
the Kerr spacetime are not fully separable in coordinate
space\footnote{However, one can separate the equations by Fourier
transforming the fields with respect to $t$ and then completing the
separation.}, this indicates that the evolution of various multipoles
are coupled.  However, Ori argues that the coupling between
multipoles should be weak at late times so that a good first
approximation to the fields can be obtained by decomposing them over
the spherical harmonics, and solving the resulting equations while
neglecting the coupling completely.  One then iterates to obtain
better approximations to the solutions.  At late times,  on a surface
of constant coordinate $r$ outside the event horizon,  the field is
given approximately by
\begin{equation}
	\Psi_{lm} \simeq F(r,\theta) e^{im\phi} t^{-(2l+2)} \; ,
\end{equation}
where $l\geq|m|$, and $F$ is
some function of both $r$ and $\theta$.  The time dependence of this
result has been verified numerically by Krivan {\it et
al}\cite{Krivan_W:1997a}.  Now, the coordinate $\phi$ is badly behaved
at the black hole event horizon. When the field is expressed in terms
of the regular coordinate $\phi_+ = \phi - \Omega_+ t$, where
$\Omega_+ = a/(2Mr_+)$, and matched to the ingoing solution at the
event horizon $r=r_+$, one arrives at
\begin{equation}
	\Psi_{lm} \simeq F_+(\theta) e^{im\phi_+} e^{im\Omega_+ v}
		v^{-(2l+2)} \; .
\end{equation}
In the final step, this ingoing solution gets matched to the solution
near to the Cauchy horizon (denoted by $r_-$) giving the final
expression
\begin{equation}
	\Psi_{lm} \simeq \left[F_v(\theta) e^{im\Omega_- v} v^{-(2l+2)} +
	F_u(\theta) e^{im\Omega_- u} u^{-(2l+2)}\right] e^{im\phi_-} \; .
\end{equation}
Here $F_u$ and $F_v$ are functions to be determined, and $\phi_-=
\phi-\Omega_- t$.  The source of the oscillations in the
non-axisymmetric modes is now obvious.

While this argument seems plausible, it is far from rigorous.  In
order to assess the complete significance of these non-axisymmetric
modes it would be necessary to establish the amplitudes of these terms
at late times, information which is currently unavailable.  For this
reason we focus our attention on the power-law decay of the wave tail;
however, we do indicate where these oscillations might modify the
analysis.

\subsection{Coordinate conditions}

While the conclusions of our analysis are couched in terms of
physical observables, such as the tidal forces (and distortion)
experienced by observers approaching the singularity inside a black
hole, or in terms of curvature scalars, it is extremely important to
understand the coordinates used to describe the spacetime.

The Cauchy horizon can be thought of as the extension of future null
infinity inside the black hole; that is, the Cauchy horizon is located
at infinite external advanced time.  Therefore it is convenient to fix
the coordinate $v$ to be standard, external advanced time as measured
by an observer far outside the black hole.  
Since the external geometry settles down to a stationary state at late
times, and the strength of the tail radiation crossing the black hole
horizon decays, we assume that $e^\lambda \sim e^{-\kappa v}$ as
$v\rightarrow \infty$; this is known to hold in non-linear evolution
of scalar fields in the spherical case, and is the coordinate
expression of the infinite gravitational blueshift between the
external universe and the Cauchy horizon.  In the subsequent analysis
we will show that this assumption leads to a self-consistent picture
of the black-hole interior.  The approach adopted here is similar in
spirit to that of BKL~\cite{BKL:1970}; we neglect terms in the field
equations which are suppressed by the exponential factor $e^\lambda$
and solve the resulting asymptotic equations.

The coordinate $u$ is chosen so that it goes to negative infinity at
the event horizon of the black hole.  In this way, it can be taken to
coincide with the natural retarded time coordinate inside a Kerr black
hole at late times.

Finally, the coordinates $\theta^a$ on the two surfaces of foliation
will be fixed as required by the analysis.

\section{A simplified model---almost plane symmetry}
 \label{sec:simple}

The goal of the present work is to examine the structure of spacetime
in the neighborhood of the Cauchy horizon of a black hole formed by
the collapse of a rotating star.  The essential features of our
analysis are most clearly illustrated in a slightly simplified context
where the Cauchy horizon is irradiated by gravitational waves of a single
polarization---this interpretation is motivated by the local ``almost
plane symmetry'' of the spacetime we consider below.

In this section we set $s_A \equiv 0$, writing the line element as
\begin{equation}
	ds^2 = -2 e^\lambda du dv + \rho^2 h_{ab} d\theta^a d\theta^b \; ,
	\label{eq:3.1}
\end{equation}
where we allow the remaining metric functions $\rho^2$, $\lambda$ and
$h_{ab}$ to depend on all the coordinates $\{u,v,\theta^a\}$.  In
addition, we assume that one may simultaneously choose coordinates in
which the conformal metric $h_{ab}$ is diagonal, and write
\begin{equation}
	h_{ab} d\theta^a d\theta^b = e^{-2\beta} dx^2 +
	e^{2\beta} dy^2 \; . \label{eq:3.2}
\end{equation}
The assumption of vanishing shift vectors reduces the Lie derivative
operators $D_A$ to simple partial derivatives, {\it i.e.} $D_A =
\partial_A$. As a result, the shear tensor has the simple form
\begin{equation}
	\sigma_{Aab}= \rho^2 \beta_{,A} {\rm diag}[-e^{-2\beta},
	e^{2\beta} ] \; , \label{shear}
\end{equation}
where a comma indicates partial differentiation. Similarly, the
expansion rates of the null rays orthogonal to the surfaces $S$ are
simply $K_A = 2 \rho^{-1} \rho_{,A}$, and the twist, given by
Eq.~(\ref{eq:2.13}), vanishes identically.

These additional assumptions reduce the Eqs.~(\ref{r1})-(\ref{r3}) to
a manageable form in which the central features of our arguments are
easily understood.  Moreover, relaxing these conditions in
Sec.~\ref{sec:general-case} results in equations which are
sufficiently similar in structure that only slight modification of the
following arguments are required to complete the general analysis.

\subsection{The asymptotic form of the equations}

In the coordinates described above, the asymptotic regime of interest
is characterized by the exponentially small value of $e^\lambda$.
This is so because we have tailored our coordinates to the physical
mechanism which underlies the Cauchy horizon instability---the
gravitational blueshift.  Moreover, the vacuum Einstein equations
(\ref{r1})--(\ref{r3}) reduce to
\begin{eqnarray}
	-D_{(A}K_{B)} - \mbox{$\textstyle \frac{1}{2}$} K_A K_B - \sigma_{Aab}
	\sigma_B\,^{ab}
	\mbox{\hspace{0.3in}}&&\nonumber \\
	+ K_{(A}D_{B)}\lambda
 	-\frac12\eta_{AB} \left(D^E+K^E\right)
		D_E\lambda &\simeq& 0 \label{eq:3.3} \\
	-\mbox{$\textstyle \frac{1}{2}$} \rho^2 h_{ab} (D_AK^A + K_AK^A)
	\mbox{\hspace{0.5in}}&&\nonumber \\
	-D^A\sigma_{Aab} + 2
		\sigma_{A(a}^{\;\;\;d} \sigma_{b)d}^A &\simeq& 0
		\label{eq:3.4}
\end{eqnarray}
and
\begin{equation}
	\sigma_{Aa;b}^{\;b} - \mbox{$\textstyle \frac{1}{2}$}\partial_a
	K_A-\frac12\partial_aD_A\lambda+\frac12K_A\partial_a\lambda
	\simeq 0 \; ,  \label{eq:3.5}
\end{equation}
where we have discarded terms in the equations which are damped by a
pre-factor $e^\lambda$.

In the following argument Eq.~(\ref{eq:3.5}) constrains the dependence of
each of the dynamical variables on $\theta^a$ near to the Cauchy horizon; it can
be ignored until later.  Contracting Eq.~(\ref{eq:3.4}) with $h^{ab}$ and
substituting for $K_A$ and $D_A$ gives
\begin{equation}
	(\rho^2)_{,uv} \simeq 0 \label{eq:3.6}
\end{equation}
which is readily solved for $\rho^2$:
\begin{equation}
	\rho^2 \simeq \rho^2_0(\theta^a) + L(v,\theta^a) + R(u,\theta^a) \; .
		\label{eq:3.7}
\end{equation}
Here $\rho_0(\theta^a)$ is assumed to be non-vanishing almost everywhere.
The functions $L(v,\theta^a)$ and $R(u,\theta^a)$ are determined by
the initial data.  The precise form depends on the detailed evolution
of the gravitational field between the event horizon and $\Sigma$; we make some
minimal assumptions below.

Substituting Eq.~(\ref{eq:3.6}) into Eq.~(\ref{eq:3.4}) provides a simple
equation for the shear, which can be expanded to a second order
equation for $\beta$ as
\begin{equation}
 	\rho \beta_{,uv} + \rho_{,u} \beta_{,v} + \rho_{,v}
		\beta_{,u} \simeq 0 \; .
        \label{eq:3.8}
\end{equation}
Defining a function $\Lambda$ by
\begin{equation}
	\Lambda =  \lambda + \ln \rho\; ,
\end{equation}
and contracting Eq.~(\ref{eq:3.3}) with $\eta_{AB}$ gives
\begin{equation}
	\Lambda_{,uv} \simeq -\sigma_{uab} \sigma{_{v}^{ab}} \; ,
    \label{Lambdauv}
\end{equation}
where the inhomogeneous term $\sigma_{uab} \sigma_v{^{ab}} = 2
\beta_{,u} \beta_{,v}$ for the metric in Eq.~(\ref{eq:3.2})

%
%
\subsection{Initial conditions}\label{subsec:simple-initial}

Initial data for Eqs.~(\ref{eq:3.3})-(\ref{eq:3.5}) is
provided on a pair of characteristic surfaces, one of which crosses
the Cauchy horizon as shown in Fig.~\ref{fig:generic-spacetime}.
This is not completely satisfactory, as it {\em assumes} the existence
of the Cauchy horizon, at least on one initial characteristic
surface.  Ideally we would like to place initial data on a spacelike
surface, such as $\Sigma$, below the inner potential barrier, but on
the past boundary of the region of high blueshift.  Unfortunately,
this program is impractical in the present context.  Instead we insure
that the solution we construct is consistent with such initial
data---the solution could be smoothly matched to that of a stationary
black hole background with small perturbations.

The choice of external advanced time $v$ to describe spacetime near
the Cauchy horizon implies that $e^\lambda \rightarrow 0$ on the
Cauchy horizon.  In particular, on the initial characteristic surface
$\Sigma^+$ ($u=u_0$) we write
\begin{equation}
	\Lambda|_+ \simeq \Lambda_0(\theta^a) - \kappa (v-v_0) + \ldots
	\label{eq:Lambda+}
\end{equation}
where $\Lambda_0(\theta^a)$ is its value on the two surface $S^0 =
\{\Sigma^+\cap \Sigma^-\}$, $\kappa$ is a constant with the dimension
of inverse length, and terms which vanish in the limit of
$v\rightarrow \infty$ are indicated by dots.  A similar condition is
demanded on $\Sigma^-$ ($v=v_0$):
\begin{equation}
	\Lambda|_- \simeq \Lambda_0(\theta^a) - \kappa (u-u_0) +
	\ldots  \; .
	\label{eq:Lambda-}
\end{equation}
Up to the surface $\Sigma$ on which effects of gravitational
blueshifting begin to take hold, the evolution of the decaying wave
tail of gravitational collapse is well described by perturbation
theory.  To the future of this surface the large blueshift suggests
that geometric optics is valid, and that ingoing gravitational waves
will be negligibly scattered by the gravitational field.  Consistent
with this picture we write the initial data for the shear as
\begin{eqnarray}
  	\beta_{,v}|_+ &\simeq& (\kappa v)^{-q/2} [ \mu(\theta^a) +
		\ldots ] \; , \label{eq:bev}\\
	\beta_{,u}|_- &\simeq& (-\kappa u)^{-p/2}[ \nu(\theta^a) +
		\ldots ] \; , \label{eq:beu}
\end{eqnarray}
where $\mu(\theta^a)$ and $\nu(\theta^a)$ are unspecified functions on
the two surfaces which foliate the two initial characteristic
hypersurfaces.  The inverse power-law decay is motivated by the
behavior of the tail radiation crossing the event horizon of the black hole.  The
precise nature of the outflux is irrelevant, although it must decay
sufficiently fast as $u\rightarrow -\infty$ \cite{Bonanno_A:1994}.
While perturbation theory suggests that $p=q$, we allow for more
general behavior.  The qualitative picture which emerges below remains
unchanged provided $\beta_{,v}|_+$ decays at least as fast as
$v^{-3/2}$ but not faster than $e^{-\kappa v}$.  In particular,
oscillatory terms which modulate the power-law decay, due to
differential rotation of the event horizon and the Cauchy horizon of a
rotating black hole, do not change the important features of the
analysis below.  These oscillatory terms will show up as the higher
order corrections in Eqs.~(\ref{eq:bev}) and (\ref{eq:beu}).

Using Eqs.~(\ref{eq:bev}) and (\ref{eq:beu}) we can now solve
Raychauduri's equation for $\rho^2$ on $\Sigma^+$ (obtained from
Eq.~(\ref{eq:3.3}) by setting $B=A=v$)
\begin{equation}
	(\rho^2)_{,vv} - (\rho^2)_{,v}\, \partial_v \Lambda =
	-2 \rho^2 (\beta_{,v})^2  \; .\label{ray}
\end{equation}
This determines the function $L(v,\theta^a)$, which can be written as
\begin{equation}
  	L(v,\theta^a)  \simeq (\kappa v)^{-q+1} \frac{2
	[\rho_0 \mu(\theta^a)]^2}{\kappa^2 (q-1)}
	\left[ 1 + O(1/v) \right]  \label{eq:L}
\end{equation}
in the large $v$ limit.  The explicit form of $R(u,\theta^a)$ is
unnecessary.  It is sufficient to note that $\partial_u R(u,\theta^a)
\rightarrow 0$ as $u\rightarrow -\infty$.

With this information, it is straightforward to check that the
curvature diverges as $v\rightarrow \infty$ along $\Sigma^+$, and that
the behavior is consistent with the discussion in Brady and
Chambers~\cite{Brady_P:1995b}.

%
%
\subsection{The evolution}

We now examine the solution determined by the initial data constructed
in the previous section.  The evolution of the gravitational degrees
of freedom (the conformal two-metric $h_{ab}$) is dictated by the
linear wave Eq.~(\ref{eq:3.8}).  Therefore the crucial step in
determining the geometry of spacetime near the Cauchy horizon is to understand the
evolution of $\beta$ from which the gravitational shear is directly
determined.  If $\beta$ was to diverge the assumptions stated in
Sec.~\ref{sec:assumptions} would be violated, and our entire
analysis would break down.

In this almost plane symmetric model, we can construct an explicit
solution of Eq.~(\ref{eq:3.8}) and demonstrate that $\beta$ remains
finite, and use it to show that $e^{-\Lambda}\rightarrow 0$ on the
Cauchy horizon for a finite range of $u$.  For the general case
presented in Sec.~\ref{sec:general-case} we are not afforded the
luxury of an exact solution, therefore we also present a method which
provides a useful bound on the shear near to the Cauchy horizon
singularity and is applicable in the general case.

The linearity of Eq.~(\ref{eq:3.8}) makes it directly amenable to
Fourier techniques (see Appendix~\ref{appendix:yurt}), however it is
more illuminating to write the solution as a series
\begin{equation}
  	\beta = \beta_0 + \rho^{-1}\sum_{i \ge 0} A_i
	\rho^{-2i} \left[ F^{(-i)}(V) + G^{(-i)}(U)
	\right] \label{SBeta}
\end{equation}
where $V=L(v,\theta^a)$, and $U=R(u,\theta^a)$.  The notation
$F^{(-i)}$ indicates the $i$-th integral of the function $F(x) \equiv
F^{(0)}(x)$ with respect to $x$.  Inserting (\ref{SBeta}) into
Eq.~(\ref{eq:beu}) and comparing integrals of the same order gives the
coefficients
\begin{equation}
	A_{i+1} = A_i \frac{(2i+1)^2}{4 (i+1)}.  \label{eq:A_recursion}
\end{equation}
The functions $F(V)$,  $G(U)$ and $A_0(\theta^a)$ are determined by
the initial data.

Differentiating Eq.~(\ref{SBeta}) with respect to $v$ and comparing
with the initial data along $\Sigma^+$ determines $F$ in the limit
$v\rightarrow\infty$ to be given by
\begin{equation}
  	F^{(1)}\, \partial_v L(v,\theta^a)
  	\simeq \frac{\rho_0\, (\kappa v)^{-q/2}}{2 A_0}
		 [\mu(\theta^a) + \ldots ] \; ,
\end{equation}
from which one can extract the behavior of $F$ to be
\begin{equation}
	F(v,\theta) \simeq \frac{\rho_0\, (\kappa v)^{1-q/2}}
	{A_0 \kappa (2 - q)}
	[\mu(\theta^a) + \ldots ] \; .
    \label{eq:F}
\end{equation}

Given the solution (\ref{eq:F}) and $L(v,\theta^a)$ in Eq.~(\ref{eq:L})
one derives the recursion relation
\begin{equation}
	F^{(-i-1)} = F^{(-i)}  \frac{(\kappa v)^{1-q}}{(i+1)(1-q)
	+1-q/2} \; .
	\label{eq:F1}
\end{equation}
Combining Eqs.~(\ref{eq:F1}),  (\ref{eq:F}) and (\ref{eq:A_recursion})
it is easy to check that the sum over $F^{(-i)}$ converges to a finite
result provided $q>3$---the boundary conditions for the radiative tail
at the event horizon actually imply $q\geq 14$.

It follows that the leading order term in Eq.~(\ref{SBeta}) is
proportional to $F$ since $F^{(-i)}(V) \gg F^{(-i-1)} (V)$ for large $v$.
This shows that $\beta_{,v}$ continues to decay with an inverse
power-law form all along the Cauchy horizon provided it has such a decay
on the initial surface.   Similar arguments provide $G$.

We can now use Eq.~(\ref{SBeta}) to compute the source term in
Eq.~(\ref{Lambdauv}) and hence determine the evolution of $\Lambda$ to
be
\begin{equation}
	\Lambda \simeq \Lambda|_-(u,\theta^a) -\kappa (v-v_0) -
	\int_{u_0}^u\! du' \int_{v_0}^v dv' \sigma_{uab}\sigma{_v}^{ab}
	\; . \label{eq:Lam}
\end{equation}
The integrand in Eq.~(\ref{eq:Lam}) is proportional to
$\beta_{,u}\beta_{,v} \sim (\kappa v)^{-q/2}$ as $v\rightarrow
\infty$, consequently the integral is finite in this limit provided
$q>2$ as we have already required.  Clearly, $\Lambda$ diverges to
negative infinity at the Cauchy horizon in precisely the same manner
as it does along the initial hypersurface $\Sigma^+$, that is
\begin{equation}
  \Lambda \sim - \kappa v
  \label{eq:La}
\end{equation}
as $v\rightarrow \infty$.  Equation~(\ref{eq:3.5}) provides a final
consistency check on the solution.

\begin{figure}
\centerline{
\psfig{file=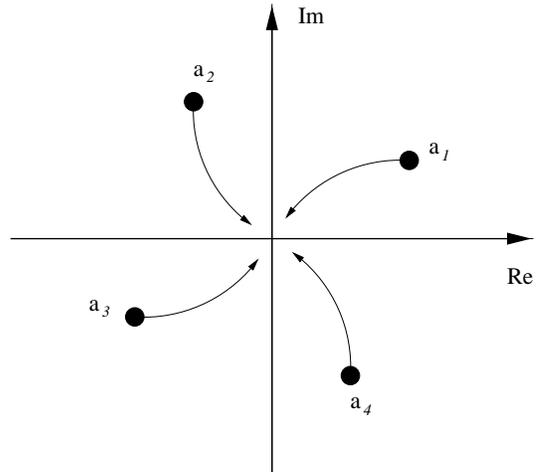,width=7cm,bbllx=124pt,bblly=232pt,bburx=488pt,bbury=560pt}
}
\caption{\label{fig:petrov}
The complex plane showing a schematic representation of the
roots of Eq.~(\protect\ref{eq:petrov-eqn}).   The arrows indicate the
evolution of the roots as $v\rightarrow\infty$.   The roots become
degenerate in the asymptotic limit,  but they do not exhibit any
peeling like properties.  This suggests that the Cauchy horizon singularity is
Petrov Type N,  and can be thought of intuitively as a shock wave
propagating into the black hole.}
\end{figure}

Having determined the approximate solution to
Eqs.~(\ref{eq:3.3})-(\ref{eq:3.5}) near to the Cauchy horizon we are in a position
to examine the curvature.  It is convenient to consider the
Newman-Penrose components of the Weyl tensor on the basis
\begin{equation}
\{ e^{-\lambda/2} \ell_{(0)}, e^{-\lambda/2} \ell_{(1)},
	e_{(a)}m^a, e_{(a)} \bar{m}^a \} \; ,
\end{equation}
where $m^a$ is a complex two-vector (called the shear axis) which
satisfies $g_{ab} = 2 m_{(a} \bar{m}_{b)}$.  The leading behavior of
each of the Weyl scalars is presented below:
\begin{eqnarray}
  \Psi_0 &&= e^{-\lambda} (\kappa v)^{-\frac{q}{2}} [ -\kappa
	\mu(\theta^a) + \ldots] \nonumber \\
  \Psi_1 &&= -
	e^{-\frac{\lambda}{2}} (\kappa v)^{-\frac{q}{2}} [ m^a
	\partial_a (\mu \nu) (-\kappa u)^{-\frac{p}{2}+1} + \ldots ]
	\nonumber \\
  \Psi_2 &&= e^{-\lambda} (\kappa v)^{-\frac{q}{2}}
	[ -\mu(\theta^a) \nu(\theta^a) (-\kappa u)^{-\frac{p}{2}}
	+\ldots ] \label{eq:Psis}\\
  \Psi_3 &&= e^{-\frac{\lambda}{2}}
	(\kappa v)^{-\frac{q}{2}+1} [\bar{m}^a \partial_a (\mu \nu)
	(-\kappa u)^{-\frac{p}{2}} + \ldots ] \nonumber \\
  \Psi_4 &&=  e^{-\lambda} (-\kappa
	u)^{-\frac{p}{2}} [ -\kappa
	\nu(\theta^a) + \ldots]. \nonumber
\end{eqnarray}
Notice that $\Psi_1$ and $\Psi_3$ are non-zero, this arises because
the spacetime is not exactly plane symmetric.  The square of the Weyl
curvature $C_{\alpha\beta\gamma\delta} C^{\alpha\beta\gamma\delta}$
diverges as $v\rightarrow \infty$ at the Cauchy horizon, and is
dominated by the radiative piece $\Psi_0 \Psi_4$ on this tetrad.
Degeneracy of the roots of the polynomial
\begin{equation}
	a^4 \Psi_4 + a^3 \Psi_3 + a^2 \Psi_2 + a \Psi_1 + \Psi_0 =0
	\label{eq:petrov-eqn}
\end{equation}
determines the Petrov classification of the
spacetime~\cite{Chandra_S:1983}.  Brady and
Chambers~\cite{Brady_P:1995b} demonstrated a four-fold degeneracy in
the limit that $v\rightarrow \infty$ on $\Sigma^+$.  Using
Eqs.~(\ref{eq:Psis}) one shows that this statement continues to hold
near to the Cauchy horizon but away from $\Sigma^+$, that is, the diverging
curvature is asymptotically Petrov type N.  This suggests the
intuitive physical picture of the singularity as a gravitational shock
wave propagating along the Cauchy horizon\footnote{This interpretation is somewhat
simplistic since the curvature scalar $C_{\alpha\beta\gamma\delta}
C^{\alpha\beta\gamma\delta}$ vanishes identically in the case of a
gravitational shock wave.}.  It is worth noting however that this
approach to type N behavior is not characterized by a peeling
property as it is at large distances outside the black hole.  Indeed
$a^2 \rightarrow \pm \sqrt{-4\Psi_0/\Psi_4}$ as $v\rightarrow\infty$
so that all four roots tend to zero.  This behavior is indicated
schematically in Fig~\ref{fig:petrov}.

\subsection{Alternative bounds}

The most important step in validating the approximation adopted in
this paper is to show that $e^\Lambda \rightarrow 0$ as $v\rightarrow
\infty$ everywhere along the Cauchy horizon. In the simple model spacetime
examined above, we have the explicit solution~(\ref{SBeta}) for the
metric function $\beta$ which determines the shear tensors through
(\ref{shear}).  Once the shear tensors are known, the evolution of
$\Lambda$ is given by the integral equation (\ref{eq:Lam}).  The
linearity of the wave equation (\ref{eq:3.8}) allows the direct
computation of $\beta$, and the explicit verification that the
integral appearing in Eq.~(\ref{eq:Lam}) vanishes in the limit
$v\rightarrow \infty$.

In a general spacetime, the evolution of the shear tensors are
described by a non-linear wave equation for which no exact solution is
known.  However, it is still possible to place bounds on the integral
appearing in Eq.~(\ref{eq:Lam}) by using the field equations, thus
verifying that $e^\Lambda \rightarrow 0$ as $v\rightarrow
\infty$. Although this procedure is unnecessary here, it is
instructive to apply it first to this case before considering the
general non-linear problem in the succeeding section.

First define a new function $\xi$ which is proportional to the
integrand,
\begin{equation}
	\xi(u,v,\theta^a) = 2 \beta_{,u} \beta_{,v} \;.
    	\label{eq:xi}
\end{equation}
Multiplying the wave Eq (\ref{eq:3.8}) for $\beta$ by $\beta_{,v}$, it
is not difficult to derive
\begin{equation}
	\partial_u \left[ \rho^2 (\beta_{,v})^2\right]
	= - \frac{1}{2} (\rho^2)_{,v} \xi.
	\label{sh1}
\end{equation}
Integrating (\ref{sh1}) with respect to $u$, making use of the
characteristic initial data (\ref{eq:bev}) and the solution
(\ref{eq:L}) for the $v$ dependence of $\rho^2$, we find\begin{equation}
(\beta_{,v})^2 = \frac{[\mu(\theta^a) \rho_0(\theta^a)]^2(\kappa v)^{-q}
} {\kappa \rho^2(u,v,\theta^a )} \left( \kappa + \int_{u_0}^{u} \xi du
\right) \label{vv}
\end{equation}
  for $u>u_0$.

Suppose that at a point with coordinates $\bar{u}$, $\bar{v}$ where
$\bar{u} \in [u_0,u]$ and $\bar{v} \in [v_0,v]$, the largest value of
the function $\xi$ occurs such that $\xi(\bar{u},\bar{v})= \bar{\xi}$.
We denote the value of any function evaluated at this point with an
overbar.  The partial derivative of $\beta$ evaluated at this point is
bounded by
\begin{equation}
\bar{\beta}^2_{,v} \le
	\frac{[\mu(\theta^a) \rho_0(\theta^a)]^2(\kappa \bar{v})^{-q} }
	 {\kappa \bar{\rho}^2} \left( \kappa + (\bar{u} - u_0)\bar{\xi}
\right).
	\label{bound1}
\end{equation}
A similar bound on the square of the partial derivative of $\beta$ with
respect
to $u$ can be derived,
\begin{equation}
\bar{\beta}^2_{,u} \leq
	\frac{[\nu(\theta^a) \rho_0(\theta^a)]^2(-\kappa \bar{u})^{-p} }
	 {\kappa \bar{\rho}^2} \left( \kappa + (\bar{v} - v_0) \bar{\xi}
\right).
	\label{bound2}
\end{equation}
It then follows from these bounds and from the definition of $\bar{\xi}$
that the inequality
\begin{eqnarray}
\bar{\xi}^2 &=& 4 \bar{\beta}^2_{,v}\bar{\beta}^2_{,u}
	 \nonumber \\
	&\le& 4 \left( \frac{\mu \nu \rho_0^2 }{\kappa \bar{\rho}^2 } 
	\right)^2
	(\kappa \bar{v})^{-q}(-\kappa \bar{u})^{-p}
	\left( \kappa + (\bar{u} - u_0) \bar{\xi} \right) \nonumber \\
        && \mbox{\hspace{1in}} \times
	\left( \kappa + (\bar{v} - v_0) \bar{\xi} \right)
 \label{ineq}
\end{eqnarray}
must be satisfied if the field equations are satisfied. This inequality
can be
rearranged into the form
\begin{equation}
	a \bar{\xi}^2 - b  \bar{\xi} -c \le 0
\label{ineq2}
\end{equation}
where the coefficients are given by
\begin{eqnarray}
	a \!&=&\! 1 - 4  \left( \frac{\mu \nu \rho_0^2 }{\kappa
		\bar{\rho}^2 } \right)^2
			(\kappa \bar{v})^{-q}(-\kappa \bar{u})^{-p}
			(\bar{u} - u_0) (\bar{v} - v_0) \label{defa} \\
	b \!&=&\! 4 \kappa  \left( \frac{\mu \nu \rho_0^2 }{\kappa 
	\bar{\rho}^2 }
		\right)^2 (\kappa \bar{v})^{-q}(-\kappa \bar{u})^{-p}
		\left( \bar{u}\! - \! u_0 \! + \! \bar{v} 
		\! - \! v_0 \right) \! \ge \! 0
		\label{defb} \\
	c \!&=&\! 4 \kappa^2 \left( \frac{\mu \nu \rho_0^2 }{\kappa 
	\bar{\rho}^2 }
	\right)^2 (\kappa \bar{v})^{-q}(-\kappa \bar{u})^{-p} \ge 0
		\label{defc}.
\end{eqnarray}
Clearly, if the coefficient $a$ in this inequality is negative, then
no bound can be placed on the maximum value of $\xi$. However, if $a$
is positive, then the upper bound on $\bar{\xi}$ is just given by
$\bar{\xi}\le b/2a +(b^2 +4ac)^{1/2}/2a$. Hence, it is important to
determine the sign of the coefficient $a$, which will depend on the
relative magnitudes of the two contributions to $a$.

Now, consider the magnitude of the second term in (\ref{defa}). At a
fixed value of $\bar{u}$, this term is proportional to $(\kappa
\bar{v})^{-q} \kappa (\bar{v} - v_0)$. In the characteristic diamond
(the region above the characteristics $\Sigma^+$ and $\Sigma^-$ in
Fig.~\ref{fig:generic-spacetime}) the advanced time coordinates
$\bar{v}$ and $v_0$ obey
\[
	\kappa \bar{v} \ge \kappa v_0  \gg 1.
\]
As a result, $(\kappa \bar{v})^{-q} \kappa  (\bar{v} - v_0) \ll 1$ in the
region of interest. The $\bar{u}$ dependence of this term is
$|\kappa \bar{u}|^{-p} \kappa (\bar{u} - u_0)$. Note that for all points
in the characteristic diamond, $|u|\le v$ and $|\bar{u}|<|u_0|$. It then
follows that
\[
	\kappa (\bar{u} - u_0) \le \kappa |u_0| \le \kappa \bar{v}
\]
so the second term of $a$ is proportional to
\begin{equation}
	\kappa (\bar{u} - u_0) \kappa (\bar{v} - v_0)
	(\kappa \bar{v})^{-q} |\kappa \bar{u}|^{-p}
	\le
	(\kappa \bar{v})^{-q+2}  |\kappa \bar{u}|^{-p}
 	\ll 1,
	\label{xx}
\end{equation}
if $q\ge3$. The prefactor $ 4 \left( \frac{\mu \nu \rho_0^2 }{\kappa
\bar{\rho}^2 } \right)^2 $ which multiplies the expression (\ref{xx})
in Eq. (\ref{defa}) is finite, and as a result, the coefficient $a$ is
positive and approximately unity.  Hence the quadratic inequality can
be used to place the following limit on $\bar{\xi}$,
\begin{equation}
 \bar{\xi} \le  	 2
	 \frac{\mu \nu \rho_0^2 }{ \bar{\rho}^2 }
	(\kappa \bar{v})^{-q/2}(-\kappa \bar{u})^{-p/2}
\label{max}
\end{equation}

A recent calculation \cite{Ori_A:1997b} suggests that the correct initial
data for $\beta$ should be an inverse power law (\ref{eq:bev}) modulated
by an oscillatory function of the form $\cos (\kappa v)$. Note that the
arguments leading to the bound (\ref{max}) will be unaltered by a
modulation of this sort, since the cosine function is bounded by one.

Given the bound (\ref{max}) on $\bar{\xi}$, we are now in a position to
integrate Eq. (\ref{eq:Lam}) and solve for the metric function $\Lambda$,
\begin{equation}
	\Lambda(u,v,\theta^a) \le \Lambda|_-(u,\theta^a)
		-\kappa(v-v_0)
		-(u-u_0)(v-v_0) \bar{\xi} .
	\label{maxLambda}
\end{equation}
Since we have shown that $\bar{\xi}$ is bounded by a vanishingly small
function, the integrated term in (\ref{maxLambda}) is negligible  compared
to the homogeneous terms which arise from the initial data. Hence,
$e^\Lambda$ vanishes at the Cauchy horizon for all $u>u_0$.

%
%

\section{The generic case}
\label{sec:general-case}

Our attention so far has focused on the toy model of the interior
presented in the previous section. Our approach was to use a metric
with only three free functions of four variables, and choose initial
data motivated by the theory of scattered fields on a stationary
background. Although the almost plane symmetric model of the previous
section does not have sufficient degrees of freedom to describe the
evolution of generic gravitational perturbations, the model is useful
since we were able to write an explicit solution and show that a weak,
null curvature singularity occurs at the spacetime's Cauchy horizon.
The importance of the toy model will become apparent in this
section where we show that the metric of a general spacetime
near to a Cauchy horizon is, to leading order, nearly identical to the
almost plane symmetric metric.

The metric used to study the generic evolution of gravitational
perturbations
is the general $2+2$ metric (\ref{eq:line-element}).  As this metric has
eight
functions, there exists the gauge freedom to set two functions to zero. We
choose to set the shift vector $s_v^a=0$. As a result, the normal Lie
derivative operator $D_v$ reduces to the partial derivative $\partial_v$.
The
general metric which we use is then,
\begin{equation}
	ds^2=-2e^\lambda
	\,du\,dv + \rho^2 \, h_{ab}(d\theta^a+s_u^a\,du)
	(d\theta^b+s_u^b\,du).
	\label{eq:general-metric}
\end{equation}
where the conformal two-metric $h_{ab}$ has unit determinant
$h=\det||h_{ab}||=1$ and hence only represents two free functions. In
the previous section we set one of these functions to zero, but here
we will consider the evolution of the general form of the conformal
two-metric.  For the present, we will not choose any particular
representation for $h_{ab}$.

The choice of coordinates and characteristic initial data are
motivated by the discussion presented in Sec.~\ref{sec:assumptions}.
As a result of our coordinate conditions, we assume that on the
initial characteristics $\Sigma^+$ and $\Sigma^-$ the form of the
function $\Lambda = \lambda +\ln \rho$ is given by equations
(\ref{eq:Lambda+}) and (\ref{eq:Lambda-}) as in the plane symmetric
spacetime. The power law initial data for the gravitational
perturbations can be set by specifying $h_{ab}$ on the initial
characteristics or equivalently, by setting the values of the shear
tensors to
\begin{eqnarray}
  	\sigma_{vab}|_+ &\simeq& \sqrt{2}(\kappa v)^{-q/2} 
	[ \mu_{ab}(\theta^a)+
		\ldots ] \; , \label{eq:sev}\\
	\sigma_{uab}|_- &\simeq& \sqrt{2}(-\kappa u)^{-p/2}[ 
	\nu_{ab}(\theta^a) +
		\ldots ] \; , \label{eq:seu}
\end{eqnarray}
where $\mu_{ab}$ and $\nu_{ab}$ are traceless two-tensors,
$h^{ab}\mu_{ab}= h^{ab} \nu_{ab}=0$ and the shear tensors are defined
by $\sigma_{Aab} = \frac{1}{2} \rho^2 D_A h_{ab}$.  As in the previous
section oscillatory terms have been neglected in the initial data
because their presence does not alter the qualitative features of the
subsequent analysis.

The analysis of the generic case is performed in the same spirit as in
the ``almost plane'' case, and follows closely the steps taken in the
last section. As before, the solution of the metric function $\rho$ on
a $u=const.$ characteristic is found by solving Raychaudhuri's
equation,
\begin{equation}
	(\rho^2)_{,vv} - (\rho^2)_{,v}\, \partial_v \Lambda =
	- \rho^2 |\sigma_{v}|^2  \; ,\label{eq:gen-ray}
\end{equation}
where we have defined the norm of a two-tensor to be 
\begin{equation}
 |\sigma_{A}|^2 = \sigma_{Aa}{^b} \sigma_{Ab}{^a} \; . \label{eq:twonorm}
\end{equation}
Given the assumptions (\ref{eq:Lambda+}) and (\ref{eq:sev}) for the
behaviors of $\Lambda$ and $\sigma_{vab}$ on $\Sigma^+$, we find
asymptotically,
\begin{equation}
	\rho^2 |_+(v,\theta^a) = \rho^2_0(\theta^a) + L(v,\theta^a) \;,
	\label{eq:rho+}
\end{equation}
where $L(v,\theta^a) $ is the same function (\ref{eq:L}) found in the
plane symmetric spacetime.

\subsection{The shift and twist vectors}

The non-vanishing shift means that the congruence of null rays
undergoes some twist. Given our gauge choice, the twist is related to
the shift by
\begin{equation}
\omega^a = - \partial_v s_u^a \;.
\label{eq:twist}
\end{equation}
The vacuum field equation $R_{va}=0$~(\ref{r3}) specifies the
behavior of the twist on a constant $u$ hypersurface:
\begin{equation}
  	\left(\partial_v + K_v - \partial_v \lambda\right) \omega^a
	= e^{\lambda} j^a
  	\label{eq:5.1}
\end{equation}
where
\begin{equation}
	j^a= 2(\sigma{{_v}^b}_{a;b} - \mbox{$\textstyle \frac{1}{2}$} \partial_a K_v - \mbox{$\textstyle \frac{1}{2}$}
	\partial_a \partial_v \lambda + \mbox{$\textstyle \frac{1}{2}$} K_v \partial_a \lambda )
\end{equation}
is independent of $\omega^a$.
We can formally integrate Eq.~(\ref{eq:5.1}) to get
\begin{equation}
	\omega^a = \frac{e^\lambda}{\rho^2} \left[
	\omega^a_0(u, \theta^a) + \int_{v_0}^v dv' \rho^2 j^a \right]\; ,
	\label{eq:5.2}
\end{equation}
where the function $\omega^a_0(u,\theta^a)$ is determined by the
initial data on $\Sigma^-$. On $\Sigma^+$, the $v$ dependence of the
functions $j^a|_+$ are known to be inverse power laws, so the integral
in (\ref{eq:5.2}) vanishes asymptotically and the twist behaves like
\begin{equation}
	\omega^a|_+ = \frac{e^\lambda}{\rho^2} 	\omega^a_0(u_0, \theta^a)
			\rightarrow \exp( -\kappa v) \; ,  \; v\rightarrow
	\infty \;.
	\label{eq:omega+}
\end{equation}
Hence the twist vector is exponentially suppressed on $\Sigma^+$.

In our problem, we still have the freedom to fix coordinates along one
hypersurface of constant $v$ \cite{Brady_P:1996}. A natural choice is
to ask that given coordinates $\theta^a$ at one point on the Cauchy
horizon, they stay constant when Lie convected by $\ell_u$ along the
Cauchy horizon. This is equivalent to the statement that $s_u^a=0$ on
the Cauchy horizon. Given this boundary condition, and a solution for
the twist, the shift vector can be found by integrating
Eq. (\ref{eq:twist}). On $\Sigma^+$ the shift vector behaves as
\begin{equation}
 s_u^a|_+ \sim \exp( -\kappa v)  \; ,  \; v\rightarrow
	\infty \;.
	\label{eq:shift+}
\end{equation}
On the initial characteristic $\Sigma^+$ the shift vector is also
exponentially suppressed.

The behavior of the shift and twist vectors on later hypersurfaces
will depend on the behavior of the functions $j^a$, and hence of the
shear and the metric functions $\rho$ and $\Lambda$ on later
hypersurfaces. In the next part of our analysis, we will assume that
the shift and twist vectors are exponentially suppressed on later
hypersurfaces, solve the resulting equations, and show that this
assumption is self-consistent. The effect of this assumption is that
to leading order in $e^\lambda$, the shift and twist vectors drop out
of the vacuum evolution equations. The field equation
$\;^{(4)}R_{ab}g^{ab} = 0 $ reduces to equation (\ref{eq:3.6}) for
$\rho$. As in the plane symmetric spacetime, the general solution for
$\rho$ is $\rho^2 = \rho^2_0 + L(v,\theta^a) + R(u,\theta^a)$. The
field equation $R_{uv}=0$ reduces to Eq.  (\ref{Lambdauv}) for
$\Lambda$. The formal solution for $\Lambda$ on later hypersurfaces is
again given by the integral equation (\ref{eq:Lam}).

\subsection{The evolution of the shear}\label{ss:bounds-generic}

The most significant difference between the general light-like
geometry and the model in section \ref{sec:simple} is the form of the
shear tensors, $\sigma_{Aab}$. In section IV we assumed that the shear
tensors have a simple diagonal form, which resulted in a propagation
equation (\ref{eq:3.8}) linear in shear. As a result, it was fairly
simple to show that regular power law initial data for the shear
evolves via the field equations to a regular solution on later
hypersurfaces which also decays as an inverse power law. In this
section we allow a general (non-diagonal) form for the shear tensors.
Although the resulting wave equation describing the evolution of the
shear is non linear the argument presented in the last section can
still be used to place a limit on possible divergences of the shear.

The evolution of the shear tensors is governed by
\begin{eqnarray}
0 &=& ^{(4)}R^a{_b} - \frac{1}{2} \delta^a{_b}^{(4)}R^d{_d}\nonumber
\\
&=& e^{-\lambda}\left( D_{A}\sigma^{A}{_a}{^b} + K_A
\sigma^{A}{_a}{^b}
	    \right) + O(1) ,
\label{shm1}
\end{eqnarray}
where we have made use of the fact that the twist and shift vectors
are exponentially suppressed in our approximation scheme. Although Eq.
(\ref{shm1}) is linear in the shear tensor, it should be noted that
$\sigma_{Aa}{^b} = \frac{1}{2} h^{ac} \partial_A h_{ac}$ which is
non-linear in the conformal two-metric. Thus Eq. (\ref{shm1}) is a
non-linear wave equation for the two-metric.

It is useful to introduce the following matrix
\begin{eqnarray}
t_{AB} &=& \frac{1}{8\pi} \left( \sigma_{A}{_a}{^b}
	\sigma_{B}{_b}{^a}
	- \frac{1}{2} \eta_{AB} \sigma_{D}{_a}{^b} \sigma^{D}{_b}{^a}
	\right) \nonumber \\
 	&=& \frac{1}{8\pi} \hbox{diag}\left( |\sigma_u|^2, |\sigma_v|^2
	\right)
 \label{shm2}
\end{eqnarray}
with positive definite entries, and the norm of a two tensor
$|\sigma_A|^2$ was defined in Eq.~(\ref{eq:twonorm}).  The divergence
of $t_{AB}$ is
\begin{equation}
\partial_{B} t_A{^B} = \frac{1}{8\pi} \left(
	 \sigma_{B}{_b}{^a} \partial_A \sigma^{A}{_a}{^b}
	 + 2 \sigma^{A}{_b}{^a} \partial_{[A} \sigma_{B]}{_a}{^b}
	 \right) .
\label{shm3}
\end{equation}
The anti-symmetric derivative of the shear occurring in the second
term of (\ref{shm3}) is related to the twist as has been shown in
appendix B of reference \cite{Brady_P:1996} and is thus
exponentially small compared to the first term of (\ref{shm3}).
Substituting the vacuum field Eq (\ref{shm1}) into (\ref{shm3}) we
find the following evolution law for the components of $t_{AB}$:
\begin{equation}
	\frac{1}{\rho^2} \partial_{B}(\rho^2 t_A{^B}) = - \frac{1}{16\pi} K_A
        \sigma_{D}{_a}{^b} \sigma^{D}{_b}{^a} + O(e^\lambda).
\label{shm4}
\end{equation}
The quantity
\begin{equation}
	\xi = -\frac{1}{2} \sigma_{Da}{^b} \sigma^{D}{_b}{^a} =
        \sigma_{ua}{^b} \sigma_{vb}{^a}.
\label{shm5}
\end{equation}
on the right hand side of this equation is the same quantity
appearing in Eq (\ref{Lambdauv}) for the metric function
$\Lambda$.  By placing bounds on $\xi$, using Eq (\ref{shm4}),
we can show that the functional form of the initial data $\Lambda|_+$  on
$\Sigma^+$ is a good approximation for the form of $\Lambda$ on later
hypersurfaces.

First note that the definition (\ref{shm5}) for $\xi$ reduces to
(\ref{eq:xi}) in the case of the plane wave metric (\ref{eq:3.1}). In
component form the field Eqs (\ref{shm4}) are
\begin{eqnarray}
\partial_u(\rho^2 |\sigma_v|^2 ) &&= - \partial_v \rho^2 \xi
\label{sh6} \\
\partial_v(\rho^2 |\sigma_u|^2 ) &&= - \partial_u \rho^2 \xi
\label{sh7}
\end{eqnarray}
which are reminiscent of Eq (\ref{sh1}). In fact, since $\xi$
satisfies the Schwartz inequality, $|\xi|^2 \le
|\sigma_{v}|^2|\sigma_{u}|^2$, the argument encapsulated in Eqs
(\ref{sh1} - \ref{maxLambda}) hold for the general double-null metric
given the assumption made in section \ref{sec:assumptions}.

\subsection{Solution of the initial value problem}

As mentioned before the dynamical degrees of freedom are encoded in
the shear tensors. Once we have fixed a gauge and know the dynamic
evolution for the shear we can in principle calculate all the metric
functions.  Initial data for the shear is supplied on $\Sigma^{\pm}$
in the form of a pair of traceless tensors $\sigma_{vab}(v,\theta^a)|_+$
and $\sigma_{uab}(u,\theta^a)|_-$ given by Eqs.~(\ref{eq:sev}) and
(\ref{eq:seu}).  Since the evolution equations for the shear are
non-linear we do not expect to find a closed form solution, however we
can construct an approximate solution along the lines of
Eq.~(\ref{SBeta}).  Write the conformal metric explicitly as
\begin{equation}
h_{ab} = \left( \begin{array}{cc}
e^{-2\beta} \cosh \gamma & \sinh \gamma \\
\sinh\gamma & e^{2\beta} \cosh \gamma
\end{array}\right) \; .
\end{equation}
The equations for the functions $\beta$ and $\gamma$ are then
\begin{eqnarray}
	[\rho \cosh (\gamma) \beta_{,v}]_{,u} + [\rho \cosh(\gamma)]_{,v}
	\beta_{,u} &\simeq& 0  \label{eq:beta-generic}\\
	(\rho \gamma_{,v})_{,u} + \rho_{,v} \gamma_{,u}
	&\simeq& 2 \sinh(2\gamma) \beta_{,v}\beta_{,u} \; ,
	\label{eq:gamma-generic}
\end{eqnarray}
where we have neglected the terms involving the shift which is
exponentially suppressed according to the arguments above.  It is
reasonable to assume that $(\rho \cosh\gamma)_{,v}$ should be
weakly dependent on $u$, so that we can integrate
Eq.~(\ref{eq:beta-generic}) for $\beta$ and obtain
\begin{equation}
	\beta \simeq \frac{B_{\rm in}(v,x^a) + B_{\rm out}(u,x^a) }{\rho\;
\cosh\gamma}
\end{equation}
Notice that this reduces to the leading term in the series solution
for $\beta$ presented in Eq.~(\ref{SBeta}) when $\gamma=0$.  Since the
shear is bounded and small according to the arguments in
subsection~\ref{ss:bounds-generic}, we further expect that the
non-linear term in Eq.~(\ref{eq:gamma-generic}) can be treated as a
source, that is, we assume that $\gamma$ is slowly varying in the
region of interest to us.  Moreover, this term involves the product
$\beta_{,v}\beta_{,u}$ which is effectively quadratic in the
luminosity of the infalling gravitational wave tail, and therefore
less important than the boundary terms.  Thus we have
\begin{equation}
	\gamma \simeq \frac{G_{\rm in}(v,x^a) + G_{\rm out}(u,x^a)}{\rho}
\end{equation}
Finally, we fix the four free functions which appear in this solution
by reference to the initial data in Eqs.~(\ref{eq:seu}) and
(\ref{eq:sev}).  The validity of these approximations has been
confirmed by Droz by numerically integrating the
equations~\cite{Droz_S:1997}.

For vanishing shifts Eqs.~(\ref{eq:3.3})-(\ref{eq:3.5}) continue to
hold as no assumptions about the form of the shear have been made in
their derivation. The analysis proceeds along the same lines as in the
previous section, except that the source to the wave-equation
(\ref{Lambdauv}) has a more complicated functional form, but it is
still small and decaying.  We therefore recover the result in
Eq.~(\ref{eq:La}) for $\Lambda$.  Similarly the function $\rho^2$ is
recovered from integrating Raychauduri's equations.

It is now straightforward to check that the asymptotic behavior of the
Weyl scalars is of the same form as in Eqs.~(\ref{eq:Psis})---the
terms inside the square brackets are different for the general case,
however the scaling in $u$ and $v$ are identical.  Hence the intuitive
picture of the singularity as a gravitational shock propagating along
the Cauchy horizon continues to be valid in the generic case.  Thus,
we have demonstrated that the generic structure of the Cauchy horizon
singularity is qualitatively captured by the almost plane symmetric
model of section~\ref{sec:simple}, and the Cauchy horizon singularity
occurs for generic perturbations (as long as the initial data is not
too singular~\cite{Droz_S:1997}).

\section{Strength of the singularity}
\label{sec:strength}

One of remarkable things about the mass-inflation singularity inside
charged, spherical black holes is that it is weak in the sense that a
coordinate system exists in which the spacetime metric is regular at
the singularity~\cite{Ori_A:1991,Ori_A:1992}.  A similar result holds
in the context of the approximate solution presented in the previous
section.  By introducing a new coordinate $V=-e^{-\kappa v}$, the
asymptotic form of the line element is
\begin{equation}
	ds^2=  \frac{2 e^{\Lambda}}{\kappa \rho V} du dV +
	\rho^2 \, h_{ab}(d\theta^a+s_u^a\,du)
	(d\theta^b+s_u^b\,du) \label{eq:regular-metric}
\end{equation}
Combining Eqs.~(\ref{eq:La}) and (\ref{eq:3.7}) with the definition of
$V$ it is clear that $^{(4)}g_{\alpha\beta}$ is bounded as
$V\rightarrow 0$, {\it i.e.} at the Cauchy horizon.

What is the relevance of this coordinate system to observations?  This
is somewhat clarified by re-stating the result; there exists a
coordinate system in which twice integrating the curvature with
respect to the new advanced time $V$ gives a finite result.  It turns
out that the proper time $\tau$ measured by an observer crossing the
Cauchy horizon satisfies $\tau \sim V$, so the tidal acceleration
experienced by this observer diverges like $\tau^{-2}(\ln|\tau|)^{-n}$
as the null singularity is approached, where $n\geq 2l+3$.  The tidal
distortion is given by twice integrating this acceleration along the
worldline of the observer, and it is finite all the way up to the
singularity.  This rough argument suggests that the regularity of the
metric, when written in terms of the coordinate $V$, indicates that
the singularity at $V=0$ is weak, in the sense that tidal distortion
of an extended object is finite there.

The interpretation of this result is somewhat unclear.  One might be
tempted to think about the singularity along the Cauchy horizon as an
``impulsive'' singularity---while an infinite force is exerted, it
acts only for a very short time.  Such a viewpoint has been adopted by
some authors and taken to indicate that a classical continuation of
spacetime beyond the Cauchy horizon singularity might
exist~\cite{Ori_A:1991,Ori_A:1992}.  Unfortunately, this point of view
seems problematic since classical physics provides no mechanism by
which to regulate the curvature once it diverges: only {\em pure}
gravitational shock waves can be confined to a thin layer in classical
General Relativity~\cite{Balbinot_R:1991}.  Indeed, we know that
quantum effects are important in the description of spacetime near the
Cauchy horizon singularity (see, for example, the discussions
in~\cite{Anderson_W:1993,Balbinot_R:1993}), and may dramatically
change the classical picture.

\section{Conclusion}
\label{sec:conclusion}

Our results strongly indicate that a wide class of initial data can
lead to the formation of a weak, null curvature singularity inside a
black hole formed by gravitational collapse.  The analysis, which is
valid at late times near to the singularity, demonstrates that the
null character of the singularity is independent of the initial data
provided the flux of radiation entering the black hole at late times
falls off more quickly that $v^{-3}$, that is the shear
$\sigma_{vab}\sigma{_{v}}^{ab}$ decays at least this fast along the
event horizon.  Moreover, our approximate solution depends on $2
\times 2$ functions ($\sigma_{uab}(u_0,v)$,
$\sigma_{vab}(u,v_0)$, ${\rm tr} \sigma_{Aab} = 0$) corresponding to
the physical degrees of freedom of the gravitational field, and we can
therefore claim that the null singularity is generic and {\em not} an
artifact of special symmetry.

{It is important to compare our results to those of previous
analyses.  Ori has investigated the singularity inside a realistic,
rotating black hole using non-linear perturbation
theory~\cite{Ori_A:1992}.  The picture of a weak, null singularity
that emerges from our work is in agreement with the results of his
analysis.  Furthermore, our results lend support to Ori's claim that the
perturbative approach captures the essential features of spacetime
structure in the neighborhood of the Cauchy horizon singularity.}

Some work remains to be done, however.  
{In spherical models,  the null Cauchy horizon singularity is a 
precursor of a strong spacelike singularity deep inside the black hole
core~\cite{Brady_P:1995a,Burko_L:Haifa}. A similar result presumably
holds for realistic rotating black holes, however little is known
about this situation.}
We have {also} seen in section~\ref{subsec:physical} that linearized
perturbations of the Kerr black hole may be modulated by terms which
oscillate infinitely many times as $v\rightarrow\infty$; it is
important to determine how significant these oscillations are for the
variation of curvature as measured by an observer approaching the
Cauchy horizon singularity.  Two approaches seem worth pursuing to
further explore this issue: (i) linear and non-linear perturbation
theory~\cite{Ori_A:1997a,Ori_A:1992} can provide an answer, in
principle, however it requires the difficult computation of the
relative amplitudes of all the terms in the perturbation series; (ii)
An alternative approach is provided by numerical techniques, similar
to those used in the spherical case~\cite{Brady_P:1995a}.  While
double null formulations of numerical relativity encounter serious
problems when caustics form along the characteristic surfaces used in
the evolution, we have in this problem a well understood regime in
which such coordinate difficulties can surely be overcome.

\section*{Acknowledgments}

We would like to thank Valeri Frolov, Amos Ori, Don Page and Eric
Poisson for helpful comments and conversations.  We are especially
grateful to Werner Israel for his suggestions, encouragement and
shared insights.  This work was supported in part by the Natural
Sciences and Engineering Research Council of Canada, and NSF Grants
AST-94-17371 and PHY-95-07740.  P.R.B. is also grateful to the Sherman
Fairchild Foundation for financial support.

\appendix
\section{Exact plane-symmetric solution} \label{appendix:yurt}

Yurtsever has proved that singularities which form in colliding plane
wave spacetimes are generically spacelike~\cite{Yurtsever_U:1988}.  In
this context, the exact solution to Eq.~(\ref{eq:3.8}) was also worked
out by Yurtsever~\cite{Yurtsever_U:1988}.  For completeness, we
present his derivation here and clarify why the isomorphism between
the internal geometries of rotating black holes and colliding plane
wave spacetimes does not imply that singularities inside black holes
are generically spacelike.  See Ref.~\cite{Ori_A:1998} for a related
argument.

Eq.~(\ref{eq:3.6}) for the conformal factor $\rho^2$ can be viewed
as an integrability condition for a coordinate transformation from the
null coordinates $(u,v)$ to new coordinates $(a,\chi)$ defined by
\begin{eqnarray}
 	a &=& \rho^2 = \rho^2_0 + L(v,\theta^a) + R(u,\theta^a) \\
	\chi &=& L(v,\theta^a) - R(u,\theta^a) \; .  \label{rc}
\end{eqnarray}
Notice that $a$ is a time coordinate since we are inside the event
horizon of the black hole.  In terms of these coordinates the wave
equation (\ref{eq:3.8}) for $\beta$ becomes
\begin{equation}
   	\beta_{,aa} - \beta_{,\chi\chi} +
	\frac{\beta_{,a}}{a} = 0 \; .\label{eq:betarr}
\end{equation}

Since Eq.~(\ref{eq:betarr}) is manifestly independent of time, we
eliminate $\chi$ using a Fourier transform.  The result is Bessel's
equation in $a$ so that the solution is
\begin{equation}
  	\beta = \beta_0 + \int\!\! dk\,  e^{i k \chi}
	\left[ c(k) Y_0(|k| a) + d(k) J_0(|k| a) \right]
  	\label{BBeta}
\end{equation}
where $J_0$ and $Y_0$ are Bessel functions of the first and second
kinds, and the two functions $c(k)$ and $d(k)$ are determined by the
initial conditions on $\Sigma$ in Fig.~\ref{fig:generic-spacetime}.
The function $Y_0(|k|a)$ diverges as $a\rightarrow 0$, but since $a$
is always larger then zero inside a black hole (at least up to early
portion of the Cauchy horizon) $\beta$ should remain regular for all
physically relevant values of $a$ and $\chi$ inside black holes.

In contrast to the black hole case, the Cauchy horizon in colliding
plane wave spacetimes is at $a=0$ and consequently the gravitational
shear generically diverges there; this causes the catastrophic
focusing of ingoing lightrays to a spacelike singularity.  Arguments
demonstrating that spacelike singularities are generic in colliding
plane wave spacetimes cannot be directly generalized to the black hole
interiors.


\end{document}